\documentclass[fleqn,usenatbib]{mnras}

\usepackage{newtxtext,newtxmath}

\usepackage[T1]{fontenc}
\usepackage{ae,aecompl}

\usepackage{graphicx}	
\usepackage{amsmath}	
\usepackage{amssymb}	

\title[Gamma-Ray Bursts 190829A and 180720B]{Radio Afterglows of Very High Energy Gamma-Ray Bursts 190829A and 180720B}

\author[L. Rhodes et al.]{
L. Rhodes$^{1, 2}$,\thanks{E-mail: lauren.rhodes@physics.ox.ac.uk}
A. J. van der Horst$^{3, 4}$,
R. Fender$^{1, 5}$,
I. M. Monageng $^{5, 6}$,
\newauthor
G. E. Anderson $^{7}$,
J. Antoniadis $^{2, 8, 9}$,
M. F. Bietenholz $^{10, 11}$,
M. B\"{o}ttcher $^{12}$,
\newauthor
J. S. Bright $^{1}$,
D. A. Green $^{13}$,
C. Kouveliotou$^{3, 4}$,
M. Kramer $^{2}$,
S. E. Motta $^{1}$,
\newauthor
R. A. M. J. Wijers $^{14}$,
D. R. A. Williams $^{1}$,
P. A. Woudt $^{5}$
\\
$^{1}$ Astrophysics, Department of Physics, University of Oxford, Keble Road, Oxford OX1 3RH, UK\\
$^{2}$ Max-Planck-Institut f\"{u}r Radioastronomie, Auf dem H\"{u}gel 69, 53121 Bonn, Germany\\
$^{3}$ Department of Physics, the George Washington University, 725 21st Street NW, Washington, DC 20052, USA\\
$^{4}$ Astronomy, Physics and Statistics Institute of Sciences (APSIS), 725 21st Street NW, Washington, DC 20052, USA\\
$^{5}$ Department of Astronomy, University of Cape Town, Private Bag X3, Rondebosch 7701, South Africa\\
$^{6}$South African Astronomical Observatory, P.O Box 9, Observatory, 7935, Cape Town, South Africa\\
$^{7}$ International Centre for Radio Astronomy Research, Curtin University, GPO Box U1987, Perth, WA 6845, Australia\\
$^{8}$ Argelander Institut f\"{u}r Astronomie, Auf dem H\"{u}gel 71, 53121, Bonn, Germany\\
$^{9}$ Institute of Astrophysics, FORTH, Dept. of Physics, University of Crete, Voutes, University Campus, GR-71003 Heraklion, Greece\\
$^{10}$ Hartebeesthoek Radio Observatory, PO Box 443, Krugersdorp, 1740, South Africa\\
$^{11}$ Department of Physics and Astronomy, York University, Toronto, M3J 1P3, Ontario, Canada\\
$^{12}$ Centre for Space Research, North-West University, Potchefstroom 2520, South Africa \\
$^{13}$  Cavendish Laboratory, 19 J. J. Thomson Ave., Cambridge, CB3 0HE, UK\\
$^{14}$ Anton Pannekoek Institute for Astronomy, University of Amsterdam, Science Park 904, 1098 XH, Amsterdam, The Netherlands}

\pubyear{2020}

\begin{document}
\label{firstpage}
\pagerange{\pageref{firstpage}--\pageref{lastpage}}
\maketitle

\begin{abstract}
We present high cadence multi-frequency radio observations of the long Gamma-Ray Burst (GRB) 190829A, which was detected at photon energies above 100\,GeV by the High Energy Stereoscopic System (H.E.S.S.). Observations with the Meer Karoo Array Telescope (MeerKAT, 1.3\,GHz), and Arcminute Microkelvin Imager - Large Array (AMI-LA, 15.5\,GHz) began one day post-burst and lasted nearly 200 days. We used complementary data from \textit{Swift} X-Ray Telescope (XRT), which ran to 100 days post-burst. We detected a likely forward shock component with both MeerKAT and XRT up to over 100 days post-burst. Conversely, the AMI-LA light curve appears to be dominated by reverse shock emission until around 70 days post-burst when the afterglow flux drops below the level of the host galaxy. We also present previously unpublished observations of the other H.E.S.S.-detected GRB, GRB 180720B from AMI-LA, which shows likely forward shock emission that fades in less than 10 days. We present a comparison between the radio emission from the three GRBs with detected very high energy (VHE) gamma-ray emission and a sensitivity-limited radio afterglow sample. GRB 190829A has the lowest isotropic radio luminosity of any GRB in our sample, but the distribution of luminosities is otherwise consistent, as expected, with the VHE GRBs being drawn from the same parent distribution as the other radio-detected long GRBs.
\end{abstract}

\begin{keywords}
radio continuum: transients, gamma-ray bursts: individual: GRB 190829A, GRB 180720B
\end{keywords}



\section{Introduction}

Gamma-ray bursts (GRBs) are caused by the launch of relativistic jets in cataclysmic events (see e.g. \citealt{Piran2004} for a review). Such bursts have gamma-ray luminosities in the range of $\sim10^{48}-10^{52}$ergs\textsuperscript{-1} and last from tens of milli-seconds to thousands of seconds \citep{1993ApJ...413L.101K, 2019ApJ...878...52A}, making them some of the most energetic transients known to date. The GRB population can be divided into two sub-groups \citep{1993ApJ...413L.101K}: short GRBs, which are thought to originate from the merger of two neutron stars \citep{1989Natur.340..126E} and have a duration less than 2\,seconds; and long GRBs, which are likely produced during the collapse of massive stars \citep{1993ApJ...405..273W} and last longer than 2 seconds.

The prompt gamma-ray emission is followed by a broadband afterglow, visible from high energy gamma-rays to radio wavelengths, which, in some cases, lasts for years (e.g. \citealt{2000ApJ...537..191F, 2008A&A...480...35V}). The prompt gamma-ray emission and subsequent afterglow are interpreted in the context of the `fireball model' \citep{1992MNRAS.258P..41R}. In the model, a relativistic blast wave propagates outwards into the circumburst medium. Initially, the outflow is highly relativistic ($\Gamma_{0} > 100$) and the prompt GRB emission is produced via processes internal to the jet \citep[see][for a review]{2000ARA&A..38..379V}. As the material in the jet interacts with the ambient medium, an external shock is produced which is observed as the afterglow. The afterglow emission originates from synchrotron cooling of an accelerated electron population with a power law distribution $N(E)dE \propto E^{-p}dE$, where p is typically between 2 and 3, and producing synchrotron emission. Such emission results in a broadband spectrum with three characteristic frequencies: the self-absorption frequency ($\nu_{SA}$); the minimum electron frequency ($\nu_{M}$), and the cooling frequency ($\nu_{C}$), which are used to derive the micro and macro GRB physics \citep{Sari1998, 1999ApJ...523..177W}. All three frequencies and the peak flux evolve with time.

The afterglow has two main components. The forward shock (FS) is produced as the jet propagates out into the circumburst medium of some density profile, usually modelled as $\rho \propto r^{-k}$. There are two most common forms of \textit{k}: $k=0$ for a uniform environment, e.g. the interstellar medium (ISM) or $k=2$ for a wind density profile \citep{Sari1998, 1999ApJ...520L..29C}. The evolution of the FS component is dependent on the density of the environment surrounding the GRB, the energy of the GRB and the fraction of the shock energy that goes into the electrons and the magnetic fields.

In some events, emission from a second component, which corresponds to the reverse shock (RS), is observed. The reverse shock propagates back towards the newly formed compact object through the ejected material \citep{Sari1999}. The synchrotron emission from the reverse shock generally appears as a declining component which  dominates at early times at optical through to radio wavelengths, but usually fades on the timescale of days as the FS becomes dominant \citep{2013ApJ...776..119L}. The  evolution of the RS is dependent on the depth of material behind the FS \textemdash either a thick or thin shell. The evolution of both the thick, where the RS becomes relativistic as it crosses the shell, and thin shell, where it remains Newtonian, varies with the surrounding density profile \citep[e.g.][]{vanderhorst2014}. In the case of the thin shell, the evolution is also dependent on whether the jet undergoes radiative or adiabatic expansion \citep{1993ApJ...405..278M}.

In some cases, the picture is not so simple: the presence of scintillation and possible `refreshed' shocks adds difficulty in discerning the overall picture \citep{scintillation, refreshed_shocks}. Scintillation occurs when emission from a compact source passes through the ISM of the Milky Way. The light curve appears to have short term, frequency dependent variations \citep{1997NewA....2..449G}. Refreshed shocks manifest as flux excesses that deviate from the fireball model. With sufficiently early radio follow up, and wide frequency coverage, it is possible to detect both the reverse and forward shock emission and distinguish clearly between the two (e.g. \citealt{2014MNRAS.440.2059A, 2019ApJ...884..121L}). 

The Fermi Large Area Telescope (LAT) has previously detected photons to energies greater than 30\,GeV from GRBs such as GRB 130427A \citep{2019ApJ...878...52A}. However, despite intensive follow up campaigns, no VHE (Very High Energy, E>100GeV) emission has been detected from any GRB by any ground based Atmospheric Cherenkov Telescopes until recently \citep[e.g.][]{2014A&A...565A..16H}. Then, in the past two years, three GRBs (GRB 180720B, 190114C and 190829A) have been detected with VHE counterparts \citep{2019Natur.575..455M, 2019ApJ...885...29F, HESS_Atel}. All three sources are at relatively low redshifts: 0.654, 0.425 and 0.0785, respectively \citep{GRB180720BZ, redshift_GCN, 2019GCN.23695....1S}, which may explain why such high energy emission was observed.

In this paper, we report on observations of two of the three VHE GRBs. In section \ref{subsection:GRB 190829A Observations}, we present an extensive radio follow-up campaign of GRB 190829A, with complementary Neil Gehrels \textit{Swift} Observatory (\textit{Swift}) X-ray Telescope (XRT) data. We discuss these observations in the context of the fireball model in section \ref{subsection: Interpretation}. In section \ref{section: GRB 180720B}, we present previously unpublished observations of GRB 180720B. Finally we compare the observations published here with those of GRB 190114C from \citet{2019arXiv191109719M, 2019Natur.575..459M}, and the AMI-LA GRB radio afterglow catalogue in section \ref{section: Comparison} \citep{Anderson2017, 2019MNRAS.486.2721B}.

\section{H.E.S.S. GRB 190829A}

\subsection{Observations}
\label{subsection:GRB 190829A Observations}

GRB 190829A was first reported by the Fermi Gamma-ray Burst Monitor (GBM) at 19:55:53\,UT (T\textsubscript{0}, \citealt{Fermi_GCN}) and shortly thereafter by the \textit{Swift} Burst Alert Telescope (BAT) on 29\textsuperscript{th} August 2019 \citep{BAT_GCN} and followed by XRT. Four hours post-detection, the High Energy Stereoscopic System (H.E.S.S.) started observing the position of the GRB. VHE emission was detected at around 20$\sigma$ significance \citep{HESS_Atel}. Spectroscopic measurements with the Gran Telescopio Canarias showed the host galaxy to be 10" from the GRB's XRT localised position and placed the host at z = 0.079$\pm$0.005, making it one of the closest GRBs detected so far \citep{redshift_GCN}.  

\subsubsection{MeerKAT}

GRB 190829A was observed with the Meer Karoo Array Telescope (MeerKAT) for ten epochs, the first starting 2.38 days post-burst, and varying between 30 and 90 minutes in duration. The observations were carried out at a central frequency of 1.28 GHz and bandwidth of 856 MHz split into 4096 channels. The primary calibrator used was J0408-6565, which was observed at the start of each observation for 5 minutes. We used scans ranging from 10 to 20 minutes on the target. The secondary calibrator used in each of the observations was J0240-2309, which was observed for 2 minutes per cycle. The data reduction was performed using \textsc{casa} \citep{casa}. We performed flagging of radio frequency interference (RFI), where the first and last 150 channels of the band were removed. Further flagging was performed using the auto-flagging algorithms \textsc{rflag} and \textsc{tfcrop}. The calibration was performed, using the flux density of the primary calibrator. We then solved for the phase-only and antenna-based delay corrections on the primary calibrator. The bandpass corrections for the primary were then applied. We solved for the complex gains on the primary and secondary, and proceeded to scale the gain corrections from the primary to the secondary and target source. Finally, we performed imaging using \textsc{wsclean} \citep{2014MNRAS.444..606O}. A full list of observations and results are given in Table \ref{tab:MeerKAT_obs}. The flux uncertainties include statistical uncertainties and a 10\,\% calibration error.

Figure \ref{fig:comparison} shows a point source from the MeerKAT image with coordinates consistent with those reported by XRT, i.e. the GRB, which is labelled `a', with a second source, labelled `b' with an arrow, 10" away, in black contours. The second source is identified as the host galaxy reported in \citet{NOT_GCN}, a 2MASS source: SDSS J025810.28-085719.2. No counterpart is found in the NRAO VLA Sky Survey (NVSS) catalogue. NVSS has a flux limit of 1\,mJy at 1.3\,GHz, a factor of 4 brighter than the second source detected in the MeerKAT observation. The 2MASS source is labelled `b' in Figure \ref{fig:comparison}.

\subsubsection{Arcminute Microkelvin Imager - Large Array}
\label{subsubsection:AMI}

The \textit{Swift}-BAT detection of GRB 190829A triggered observations with the Arcminute Microkelvin Imager - Large Array (AMI-LA) as part of a GRB follow up program: ALARRM \citep[AMI-LA Rapid Response Mode,][]{2013MNRAS.428.3114S, Anderson2017}.

Observations with AMI-LA commenced 1.33 days post-burst. The observation lasted three hours, using seven out of eight antennas. The data were reduced using a custom pipeline \textsc{reduce\_dc} \citep{2013MNRAS.429.3330P}. The data were flagged for RFI and calibrated using 3C286 and J0301+0118 as the bandpass and phase calibrators, respectively. We imaged and deconvolved interactively in \textsc{CASA} using the task \textsc{clean}. 

The first observation showed a 4\,mJy/beam point source, its coordinates consistent with those from XRT.  

\begin{figure}
    \centering
    \includegraphics[width=\columnwidth, trim={2.2cm 12cm 3cm 3cm}, clip]{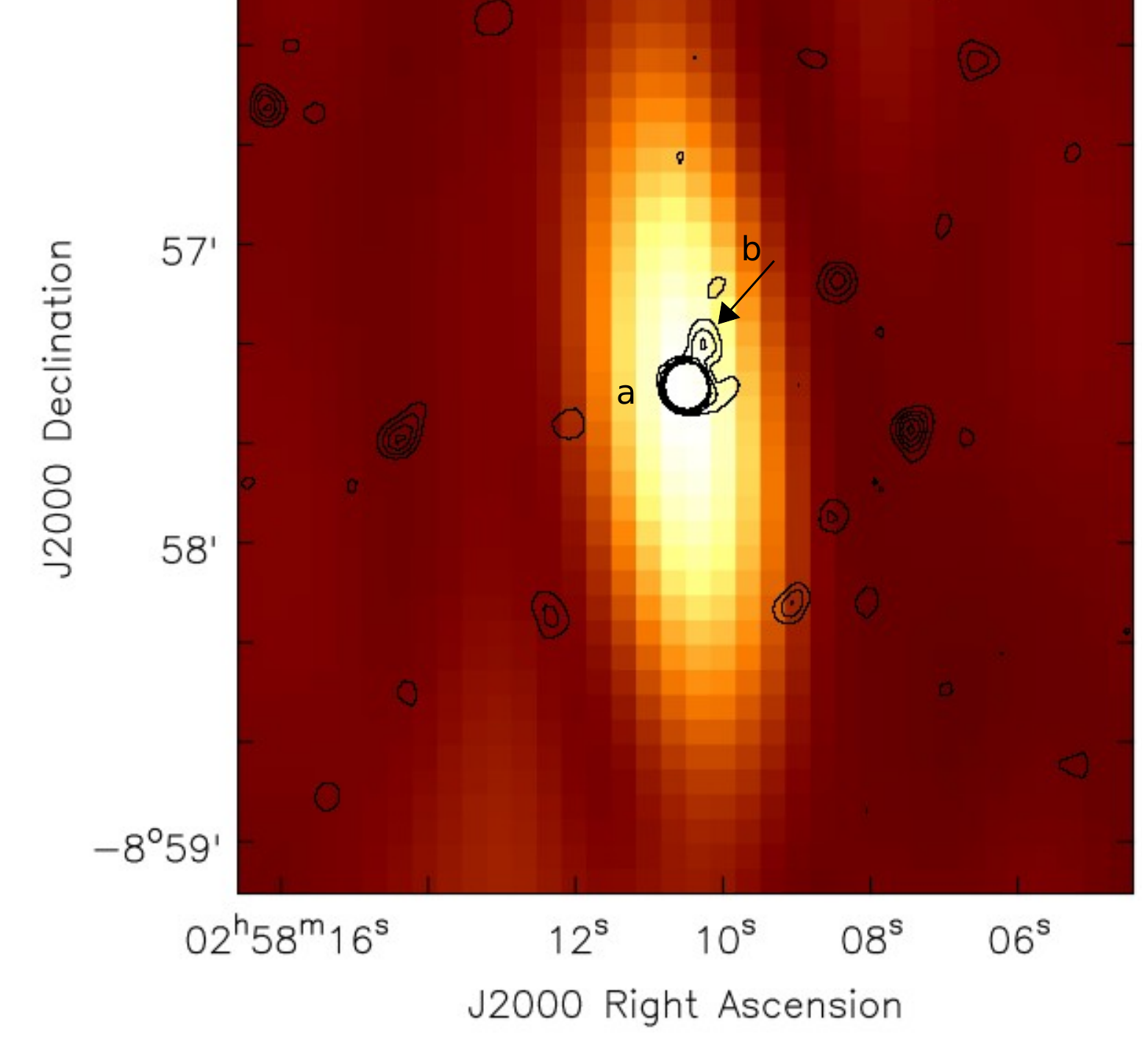}
\caption{Comparison between MeerKAT (1.3\,GHz) and AMI-LA (15.5\,GHz) images of the field of GRB 190829A. The MeerKAT image (contours) is from a concatenated measurement set of all 10 observations. The contours are at 3, 6, 9, 12, and 15$\sigma$, with an image rms noise of 8$\mu$Jy/beam. The bright object at the centre of the image, labelled `a' is the GRB and the second source seen to the north-west labelled `b' is the host galaxy. The AMI-LA (pixel) map is made from a concatenation of 5 separate observations and has a rms noise of $\sim$50$\mu$Jy/beam. In the AMI-LA observations, we do not resolve the GRB from the host galaxy (due to much shorter baselines, despite the higher frequency). This is an important factor to consider when examining the AMI-LA light curve which shows a plateau component that we believe originates from the host galaxy.}
    \label{fig:comparison}
\end{figure}

The initial detection triggered a long term monitoring campaign, with daily observations until 41 days post-burst before moving to bi-weekly observations. Observations ceased 143 days post-burst. A full list of the observations is given in Table \ref{tab:AMI_obs}. The flux uncertainties include statistical and a 5\,\% calibration error. The background colour map in Figure \ref{fig:comparison} shows a concatenation of 5 AMI-LA data sets. The AMI-LA beam was 95" by 26", the elongation due to the low declination of the source with respect AMI-LA's observing range.

\subsubsection{\textit{Swift}-XRT}

\textit{Swift}-XRT started observing the field of GRB 190829A 110s after the initial trigger (\citealt{2004ApJ...611.1005G, XRT_GCN}). XRT observed in the band 0.3-10\,keV, until 115 days after the burst. The data used in this work were collected in photon counting mode after 4000s and extracted using the \textit{Swift} Burst Analyser (\citealt{Evans2007,Evans2009}).

\subsection{Results}
\label{GRB 190829A Results }

\begin{figure*}
    \centering
    \includegraphics[width = 0.8\textwidth]{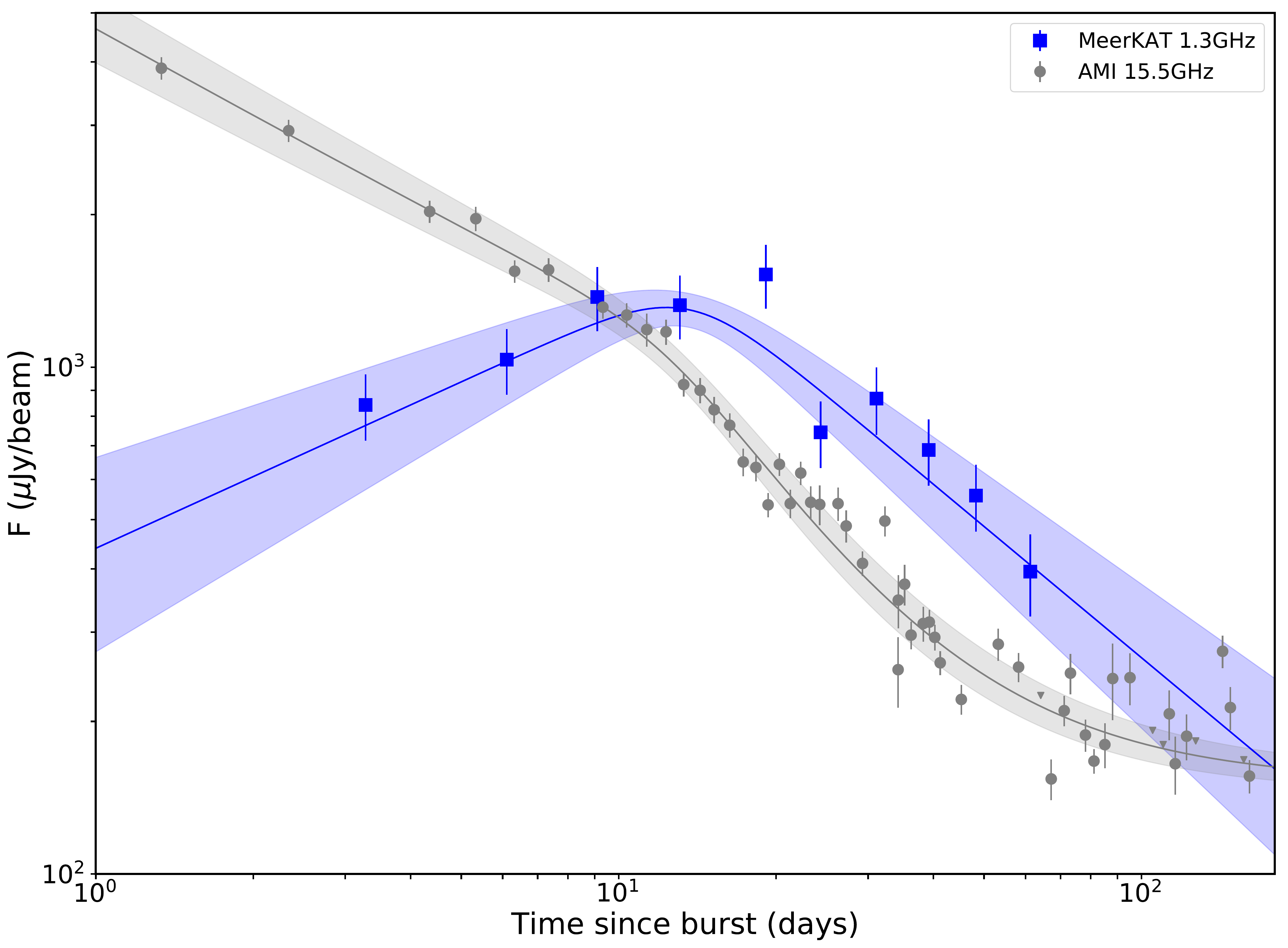}
    \caption{Light curves of the AMI-LA and MeerKAT data at 1.3\,GHz and 15.5\,GHz up to day 180, respectively, with equation \ref{eq:bpl} fit to each data set. The error bars on the data points include the statistical 1$\sigma$ uncertainty and a calibration error (5\,\% for AMI-LA and 10\,\% for MeerKAT) added in quadrature. The shaded regions represent the 16\textsuperscript{th} and 84\textsuperscript{th} percentiles from their respective fits. }
    \label{fig:lc_s}
\end{figure*}

\begin{figure}
    \centering
    \includegraphics[width = \columnwidth]{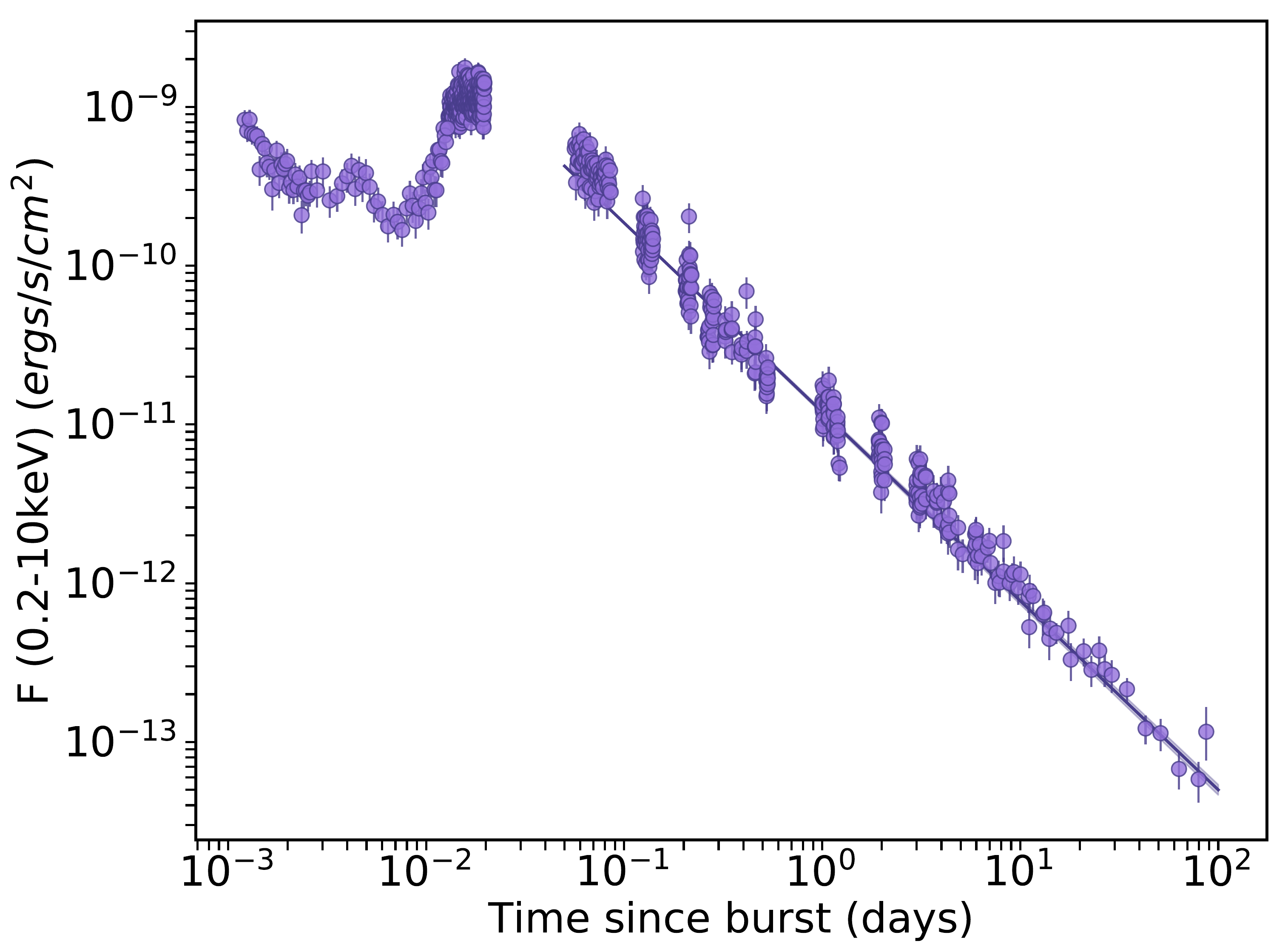}
    \caption{\textit{Swift}-XRT light curve for GRB 190829A with data points from both photon counting and window timing modes, with uncertainties quoted to 1$\sigma$. Fitted to the data, from 0.1\,days, is a power law where $F \propto t^{\alpha_{X}}$ and $\alpha_{X} = -1.19\pm0.01$.}
    \label{fig:XRT}
\end{figure}

The results of the observations described above are given in tables \ref{tab:MeerKAT_obs} and \ref{tab:AMI_obs}, and shown in Figures \ref{fig:lc_s} and \ref{fig:XRT} with the fluxes reported in Tables \ref{tab:MeerKAT_obs} and \ref{tab:AMI_obs}. We model the data with power law components of the form $F_{\nu} \propto t^{\alpha}\nu^{\beta}$, where $t$ is the time elapsed since the burst, $\nu$ refers to the central frequency of the observing band and $\alpha$ and $\beta$ are the power law indices. Subscripts refer to the frequency or energy band. 

All fits to the data were performed using \textsc{emcee} \citep{emcee}, a MCMC (Monte Carlo Markov Chain) sampler. We used 700 independent walkers, burned the first 5000 of 10000 steps, which resulted in 3,500,000 samples. Non-informative priors were used for all parameters. We use a maximum likelihood analysis to find the optimum chi-squared corresponding to the best fit. We quote the 50\textsuperscript{th} percentile of the samples in the marginalised distributions as the best fit with the 16\textsuperscript{th} and 84\textsuperscript{th} percentiles quoted as the lower and upper uncertainties, respectively.

\subsubsection{Radio}

The 1.3\,GHz MeerKAT data (the blue squares in Figure \ref{fig:lc_s}) show a rise until a maximum at around day 15 when the light curve turns over into a decay that continues for the rest of the observations. We fit the data using a smoothly broken power law equation: 
\begin{equation}
        F_{\nu}(t) = A\left[\frac{1}{2}\left(\frac{t}{t_{b}}\right)^{-5\alpha_{1}} + \frac{1}{2}\left(\frac{t}{t_{b}}\right)^{-5\alpha_{2}}\right]^{-\frac{1}{5}} + B
    \label{eq:bpl} 
\end{equation}

The free parameters in the fit were A, t\textsubscript{b}, $\alpha_{1}$, $\alpha_{2}$ and B; where A is the amplitude, t\textsubscript{b} is the break time, $\alpha_{1}$ and $\alpha_{2}$ are the exponents of the two power laws, and B is a constant offset. The factor of 5 is a fixed smoothness parameter. 

For the MeerKAT data set, B was set to zero. We used priors of: 500 < A\,($\mu$Jy/beam) < 3500, 0 < $\alpha_{1}$ < 3, -3 < $\alpha_{2}$ < 0, and 10 < t\textsubscript{b}\,(days) < 30. The results for the 1.3\,GHz fit are shown in Table \ref{tab:fit_results} with the corner plot given in Appendix \ref{fig:corner MeerKAT}.

\begin{table}
 \caption{List of observations made with MeerKAT at 1.3\,GHz. Each with the time since burst (T\textsubscript{0}), the flux density with uncertainties (including statistical and 10\,\% calibration error) and duration.}
 \label{tab:MeerKAT_obs}
 \begin{center}
 \begin{tabular}{ccc}
  \hline
   T-T\textsubscript{0}(days) & Flux ($\mu$Jy/beam) & Duration (hrs) \\
  \hline
  3.28&840$\pm$130&1.33\\
  6.11&1040$\pm$150&0.72\\
  9.10&1380$\pm$200&0.72\\
  13.10&1330$\pm$190&0.72\\
  19.13&1520$\pm$220&0.27\\
  24.34&740$\pm$110&0.47\\
  31.12&870$\pm$130&0.27\\
  39.18&690$\pm$100&0.57\\
  48.26&560$\pm$80&0.38\\
  61.29&400$\pm$70&0.42\\
  \hline
 \end{tabular}
 \end{center}
\end{table}

The AMI-LA data (grey circles in Figure \ref{fig:lc_s}) are best described by two decaying power laws with a break around day 12 followed by a plateau component after 70 days. The plateau component, most likely, can be attributed to the host galaxy because of the second component seen in the MeerKAT field within the AMI-LA beam in Figure \ref{fig:comparison}. The galaxy component has an 1.0-1.5\,GHz in-band spectral index of $\beta_{1.0-1.5} = 0.5\pm0.7$. Extrapolating to 15.5\,GHz, the flux density of the source would be $10^{-(3.6\pm1.6)}$Jy/beam. The flattening of the light curve seen from around day 70 has a flux level within the uncertainties of the predicted host galaxy flux at 15.5\,GHz. 

The AMI-LA light curve shows day-to-day variability which increases in amplitude in the plateau section of the light curve with fractional variability $\sim15\%$. We cannot attribute the variability to scintillation because it is not seen in the earliest epochs. If due to scintillation, we would observe flux variation at the earliest times when the jet is most compact. The variability could be due to a combination of telescope pointing error, intrinsic variability and seeing. However, we do not consider this variability to be real. Equation \ref{eq:bpl} was fit to the data, with flat priors of: 0 < A\,($\mu$Jy/beam) < 5000, -1 < $\alpha_{1}$ < 0, -3 < $\alpha_{2}$ < -1, 5 < t\textsubscript{b}\,(days) < 20 and 0 < B\,($\mu$Jy/beam) < 500. A list of fluxes are given in Table \ref{tab:AMI_obs} and the results of the fits are given in Table \ref{tab:fit_results} with the associated corner plot given in Appendix \ref{fig:corner AMI}.

\begin{table}
 \caption{List of observations made using AMI-LA at 15.5\,GHz. Each with  the time since burst (T\textsubscript{0}), the flux density and uncertainties (including statistical and 5\,\% calibration error) and duration. On occasions where the source was not detected we provide a 3 $\sigma$ upper limit with the prefix `<'.}
 \label{tab:AMI_obs}
 \begin{center}
 \begin{tabular}{ccc}
  \hline
  T-T\textsubscript{0}(days) & Flux ($\mu$Jy/beam) & Duration (hrs) \\
  \hline
  1.34&3890$\pm$200&3\\
  2.34&2930$\pm$150&3\\
  4.35&2030$\pm$100&2\\
  5.33&1960$\pm$110&3\\
  6.33&1550$\pm$80&3\\
  7.35&1560$\pm$80&3\\
  9.33&1310$\pm$70&4\\
  10.36&1270$\pm$70&2\\
  11.32&1190$\pm$90&2\\
  12.32&1170$\pm$70&4\\
  13.32&920$\pm$50&4\\
  14.32&900$\pm$50&4\\
  15.23&830$\pm$50&4\\
  16.31&770$\pm$40&4\\
  17.31&650$\pm$40&4\\
  18.31&630$\pm$40&4\\
  19.32&540$\pm$30&3.3\\
  20.30&640$\pm$30&4\\
  21.30&540$\pm$40&4\\
  22.30&620$\pm$30&4\\
  23.30&540$\pm$40&3.5\\
  24.25&540$\pm$50&2\\
  27.24&550$\pm$40&1.5\\
  28.28&490$\pm$40&4\\
  29.28&410$\pm$20&2\\
  32.30&500$\pm$30&4\\
  34.27&350$\pm$40&4\\
  34.24&250$\pm$40&2\\
  35.24&370$\pm$30&2\\
  36.26&300$\pm$20&3\\
  38.25&310$\pm$30&4\\
  39.29&310$\pm$20&4\\
  40.25&290$\pm$20&4\\
  41.22&260$\pm$10&2\\
  45.23&220$\pm$20&4\\
  53.23&280$\pm$20&3\\
  58.21&260$\pm$20&4\\
  64.18&$<$230&4\\
  67.21&150$\pm$10&3\\
  71.17&210$\pm$20&3\\
  73.16&250$\pm$20&4\\
  78.14&190$\pm$10&3\\
  81.13&170$\pm$10&4\\
  85.12&180$\pm$20&4\\
  88.11&230$\pm$40&4\\
  95.10&240$\pm$30&4\\
  105.07&$<$190&4\\
  110.06&$<$180&4\\
  113.05&210$\pm$20&4\\
  116.04&170$\pm$20&4\\
  122.02&190$\pm$20&4\\
  127.01&$<$180&4\\
  142.96&280$\pm$20&4\\
  \hline
 \end{tabular}
 \end{center}
\end{table}

\begin{table*}
 \caption{The results of the MCMC fitting code parameters which describes the MeerKAT 1.3\,GHz and AMI-LA 15.5\,GHz light curves as given by equation 
 \ref{eq:bpl} and the X-ray light curve by a single power law function. The values quoted are the mean with uncertainties at the 16\textsuperscript{th} and 84\textsuperscript{th} percentile.}
 \label{tab:fit_results}
 \begin{center}
 \begin{tabular}{cccccc}
  \hline
  Observing range &  A\,($\mu$Jy/beam) & $\alpha_1$ & $\alpha_2$ & $t_{\textrm{b}}$\,(days) & B\,($\mu$Jy/beam) \\
  \hline
  1.3\,GHz & $1400\pm100$ & $0.5\pm0.1$ & $-0.9\pm0.1$ & $14\pm2$ &  $-$ \\
  15.5\,GHz & $880^{+80}_{-70}$ & $-0.59\pm0.03$ & $-1.71^{+0.08}_{-0.09}$ & $12.6\pm0.9$ & $152^{+7}_{-8}$ \\
  0.2-10keV & $0.50\pm0.01$ & $-1.19\pm0.01$ & $-$ & $-$ & $-$ \\
  \hline
 \end{tabular}
 \end{center}
\end{table*}

\subsubsection{X-Rays}

The full XRT light curve is given in Figure \ref{fig:XRT}. The early time \textit{Swift}-XRT data are highly variable due to the prompt emission that originates from the GRB itself and not the afterglow. After this initial phase, there is a decaying flux component from $\sim$10\textsuperscript{-1} days. This component can be described by a single power law: $F \propto t^{\alpha_{X}}$, where $\alpha_{X} = -1.19\pm0.01$.

We also examine the spectral properties of the XRT data. The late time-averaged spectrum (after 4400s), from the \textit{Swift} Burst Analyser \citep{Evans2010}\footnote{https://www.swift.ac.uk/xrt\_spectra/00922968/} is characterised by a photon index of $\Gamma = 2.10\pm0.09$, corresponding to a spectral index of $\beta_{0.2-10\textrm{keV}} = -1.01\pm0.09$.

\subsection{Interpretation}
\label{subsection: Interpretation}

Here we present the results of our observations in the context of the fireball model in which a shock propagating forward into the surrounding medium accelerates electrons producing a time-evolving synchrotron spectrum.

\subsubsection{X-Rays: Forward Shock}

The late-time \textit{Swift}-XRT light curve shows a single power law decline with no breaks, indicating that no break frequency passes through the observing band. If $\nu_{C}$ passed through the band we would expect to see the slope of the light curve steepen. The data can be described by a power law in the form of $F\propto t^{-1.19\pm0.01}$, and is steep enough in time to be above the cooling break of a FS component giving $p = 2.25\pm0.02$ (where p, is the power law exponent from the electron energy distribution), independent of the GRB's surrounding density profile \citep{Granot2002}. The late-time-averaged spectrum also shows that the observed emission originates above the cooling break. A photon index of $\Gamma = 2.10\pm0.09$ equates to a spectral index of $\alpha = -1.10\pm0.09$ and for the X-ray regime above the cooling break we get p = $2.2\pm0.2$. 

\subsubsection{Radio: Forward Shock}

The 1.3\,GHz MeerKAT observations also fit with the FS model. The rise in flux up to day $\sim$15 is consistent, within errors, with emission above $\nu_{SA}$ and below $\nu_{M}$ as the FS propagates through a homogeneous environment. The model, from e.g. \citet{Granot2002}, gives $F \propto t^{0.4}$ for this frequency range, which is in agreement with our fitted value of $0.5\pm0.1$.

The turnover at 15 days at 1.3\,GHz corresponds to the peak frequency ($\nu_{M}$) passing through the observing band. The decay component follows $F\propto t^{-0.71^{+0.08}_{-0.09}}$. Using the scaling for a homogeneous medium, where the exponent is equal to $\frac{3(1-p)}{4}$, results in $p = 2.1\pm0.3$, a result consistent with that from the XRT light curve and time averaged spectra. 

There is a difference of $0.3\pm0.1$ between the exponents of the XRT and post-break MeerKAT light curves, with the MeerKAT one being the shallower of the two. Such a difference indicates that $\nu_{C}$ could be between the two observing bands. The theoretical difference in the temporal power law decay exponent caused by $\nu_{C}$ moving between the bands is $\Delta \alpha = 0.25$. This is fully consistent with the results seen in the MeerKAT and XRT data, indicating that $1.3\,\textrm{GHz} \leq \nu_{C} \leq 0.2$\,keV.

The AMI-LA data set is inconsistent with the FS model. On top of the host galaxy emission, the early decay follows $F\propto t^{-0.59\pm0.03}$. If the emission originates from the optically thin part (i.e. above $\nu_{M}$) of the FS, the light curve would give $p = 1.80\pm0.09$ . The early 15.5\,GHz decay is too shallow to be consistent with the XRT light curve slopes. Furthermore, if the MeerKAT and AMI-LA light curves are both from the FS, we would expect to see a break in the AMI-LA data at an earlier time as a result of $\nu_{M}$ passing through the band. We can use the break time at 1.3\,GHz to calculate the expected break time at 15.5\,GHz. The frequency break $\nu_{M}$ evolves as $t^{-\frac{3}{2}}$ independent of the structure of the surrounding environment; working backwards one expects to see a break at about 2.7$\pm$0.4 days in the AMI-LA light curve. No such break is observed and the observed flux levels are too high to be consistent with the FS seen in the MeerKAT data.

\begin{figure}
    \centering
    \includegraphics[width=\columnwidth]{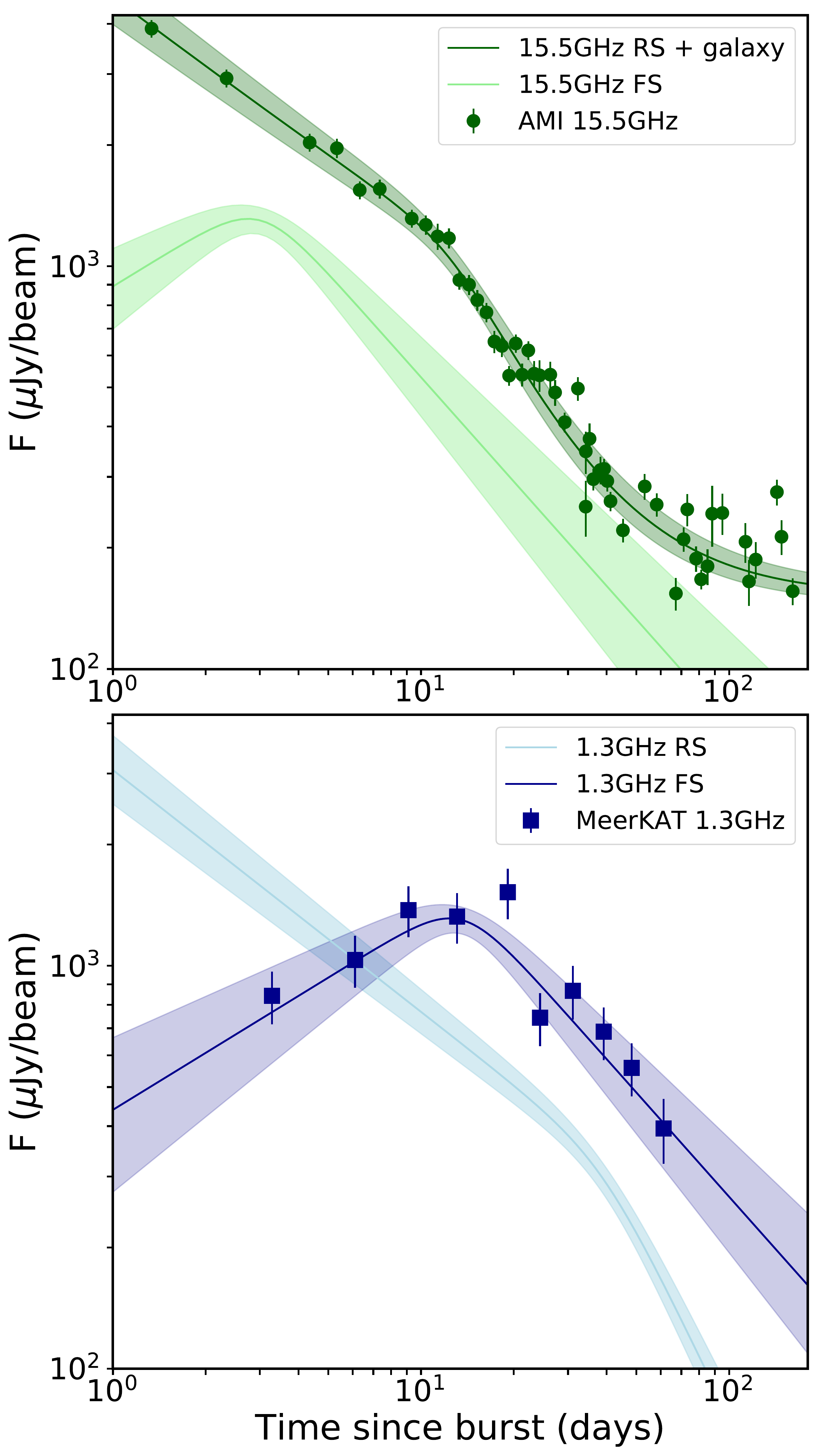}
    \caption{\textit{Upper panel:} The AMI-LA light curve, in dark green, shows the RS and underlying galaxy component. The light green broken power law shows how the FS would evolve in a homogeneous environment at 15.5\,GHz. \textit{Lower panel:} The FS-dominated MeerKAT light curve with the theorised RS component as seen at 1.3\,GHz, derived from the 15.5\,GHz light curve.}
    \label{fig:RS_FS}
\end{figure}

The AMI post-break decay, if due to optically thin emission (i.e. $\nu_{M} \leq 15.5\,\textrm{GHz} \leq \nu_{C}$), with an exponent of $-1.71^{+0.08}_{-0.09}$ gives $p = 3.3\pm0.2$, for a homogeneous medium, a value which is inconsistent with that from the X-ray light curve. The slope of the light curve is too steep to originate from a FS component. 

We can also rule out the possibility of a jet break. A jet break would result in a simultaneous steepening of both radio light curves to a power law relation of $F\propto t^{-p}$. While we see breaks in the AMI and MeerKAT light curves at the same time, the post break decays both too shallow to be from a jet break and are inconsistent with each other. Both slopes are too shallow to originate from a jet break in the FS.

\subsubsection{Radio: Forward and Reverse Shock}

The AMI light curve can be explained when we include an additional component in our model: a RS. We consider both the thick and thin shell regimes; the pre-break emission in the AMI light curve is the most similar to the RS where $\nu_{SA} \leq 15.5\,\textrm{GHz} \leq \nu_{M}\textrm{,} \nu_{C}$. In the thick shell regime, for $k = 0$ the power law exponent of the slope would be -0.47 and for $k = 2$, it would be -0.5. In the thin shell regime, the expected slope is around -0.46 \citep{vanderhorst2014}. These are all shallower than, but close to, the fitted result of $0.59\pm0.3$.

Post-break, the slope falls at a rate consistent with optically thin synchrotron ($\nu_{SA}  \leq \nu_{M} \leq 15.5\,\textrm{GHz} \leq \nu_{C}$) from the RS, for a homogeneous environment and $p = 2.5\pm0.1$ in a thick shell regime. The decay is too steep to be physically representative in the thin shell regime. 

We can check if the FS contributes significantly to the AMI-LA light curve. Using a spectral index of $-0.75\pm0.03$, derived from the MeerKAT optically thin data, we extrapolate the 1.3\,GHz light curve to 15.5\,GHz. Using the 1.3\,GHz data points and table 1 from \citet{vanderhorst2014}, we produced a theoretical FS light curve at 15.5\,GHz. The simulated light curve is shown in the upper panel of Figure \ref{fig:RS_FS} along side the original AMI data set with their respective fits and shaded 68\% confidence level uncertainty regions. At all times the FS shock emission at 15.5\,GHz is fainter than the flux values we measure. Past 30\,days, the simulated FS component becomes comparable to the AMI light curve but this is also where emission from the host galaxy begins to dominate. 

Similarly, we can check if the RS component detected at 15.5\,GHz contributes significantly to the 1.3\,GHz MeerKAT light curve. As such, we extrapolate our early time AMI data points, using a spectral slope of $F \propto \nu^{\frac{1}{3}}$ (we assume both 1.3 and 15.5\,GHz are between $\nu_{SA}  \textrm{ and } \nu_{M}$), and plot the predicted RS light curve  at 1.3\,GHz in the lower panel of Figure \ref{fig:RS_FS}. Except for the first data point, where the theoretical 1.3\,GHz RS dominates over the FS, it is clear that the FS is the main emission component in the MeerKAT data set. This early emission at 1.3\,GHz could be suppressed by synchrotron self-absorption of RS emission by the FS.

In summary, we interpret our radio and X-ray observations of GRB 190829A as a combination of two shocks: MeerKAT and XRT light curves show forward shock emission and the AMI-LA light curve shows a reverse shock component fading until around 70 days post burst where we see emission from the host galaxy. 

\section{H.E.S.S. GRB 180720B}
\label{section: GRB 180720B}

GRB 180720B was detected by \textit{Swift}-BAT on 2018 July 20 14:21:44\,UT, \citep{2018GCN.22973....1S}. X-Shooter VLT observations placed the GRB at a redshift of 0.654 \citep{GRB180720BZ}. H.E.S.S identified a 5$\sigma$ source consistent with the GRB's position 10 hours after the initial detection report \citep{2019Natur.575..464A}.

\subsection{Observations: AMI-LA}

The \textit{Swift}-BAT detection of GRB 180720B triggered observations with AMI-LA  \citep{2013MNRAS.428.3114S, Anderson2017}. In total, 5 logarithmically spaced observations were made. The observations were reduced using the same method described in section \ref{subsubsection:AMI}. The list of fluxes and upper limits measured are listed in Table \ref{tab:AMI_obs_GRB_180720B}. Our results are plotted in a lower panel of Figure \ref{fig:GRB_190114C}.

\subsection{Results and Interpretation}

\begin{table}
 \caption{Peak fluxes and 3$\sigma$ upper limits of 15.5\,GHz observations for GRB 180720B. Observations were made with AMI-LA as part of the ALARRM \citep{2013MNRAS.428.3114S, Anderson2017}. Here, we give the time since burst (T\textsubscript{0}), the peak flux for each epoch with a detection along with uncertainties (including statistical and 5\% calibration error) and duration. On occasions where the source was not detected, we provide a 3$\sigma$ upper limit with the prefix `$<$'.}
 \label{tab:AMI_obs_GRB_180720B}
 \begin{center}
 \begin{tabular}{ccc}
  \hline
   T-T\textsubscript{0}(days) & Flux ($\mu$Jy/beam) & Duration (hrs) \\
  \hline
    1.69	&	1100$\pm$60&4	\\
    3.66	&	580$\pm$50&2	\\
    5.65	&	340$\pm$40&4	\\
    6.66	&	<220	&	4.5	\\
    25.59	&	<190	&3		\\
  \hline
 \end{tabular}
 \end{center}
\end{table}

We characterise the emission using a steep power law decay which is seen up until 6 days after which only 3$\sigma$ upper limits of around 200\,$\mu$Jy/beam, were obtained. Using a power law fit, the decay follows $F\propto t^{-1.2\pm0.1}$. Such a fit is consistent with optically thin emission ($\nu_{M} \leq 15.5\,\textrm{GHz} \leq \nu_{C}$) from the FS
for both a ISM and stellar wind profile. However, the light curve is also steep enough to originate from emission above the cooling break ($15.5\,\textrm{GHz} \leq \nu_{C}$). 

To resolve the degeneracy of which spectral branch of the FS the AMI-LA emission comes from, we compared the AMI-LA data to the published XRT data from \citet{2019ApJ...885...29F} and the optical data from \citet{GRB180720B}. The XRT light curve for this event shows a power law decay between days 0.02 and $\sim$3 of $F_{\textrm{XRT}}\propto t^{-1.26\pm0.06}$, which they also attribute to optically thin synchrotron emission. A break is seen at 3 days after which the light curve decays at a steeper rate, due to the cooling break passing through the XRT band. The optical data, up until day 12, are ascribed to optically thin synchrotron emission from a forward shock component ($F_{\textrm{opt}}\propto t^{-1.22\pm0.2}$) in a homogeneous medium. If these are correct, the radio emission is also likely due to the optically thin emission from the FS given the very similar temporal slope.

\section{Comparison with other GRBs}
\label{section: Comparison}

There are now three GRBs that have had VHE emission detected by Cherenkov arrays. Here, we compare the radio data for the three GRBs in two ways: examining the measured flux values at different bands and comparing the spectral luminosities at around 16\,GHz to a sensitivity-limited selection of the GRB population.

\subsection{Radio light curves of the VHE GRBs}

\begin{figure}
    \centering
    \includegraphics[width=\columnwidth]{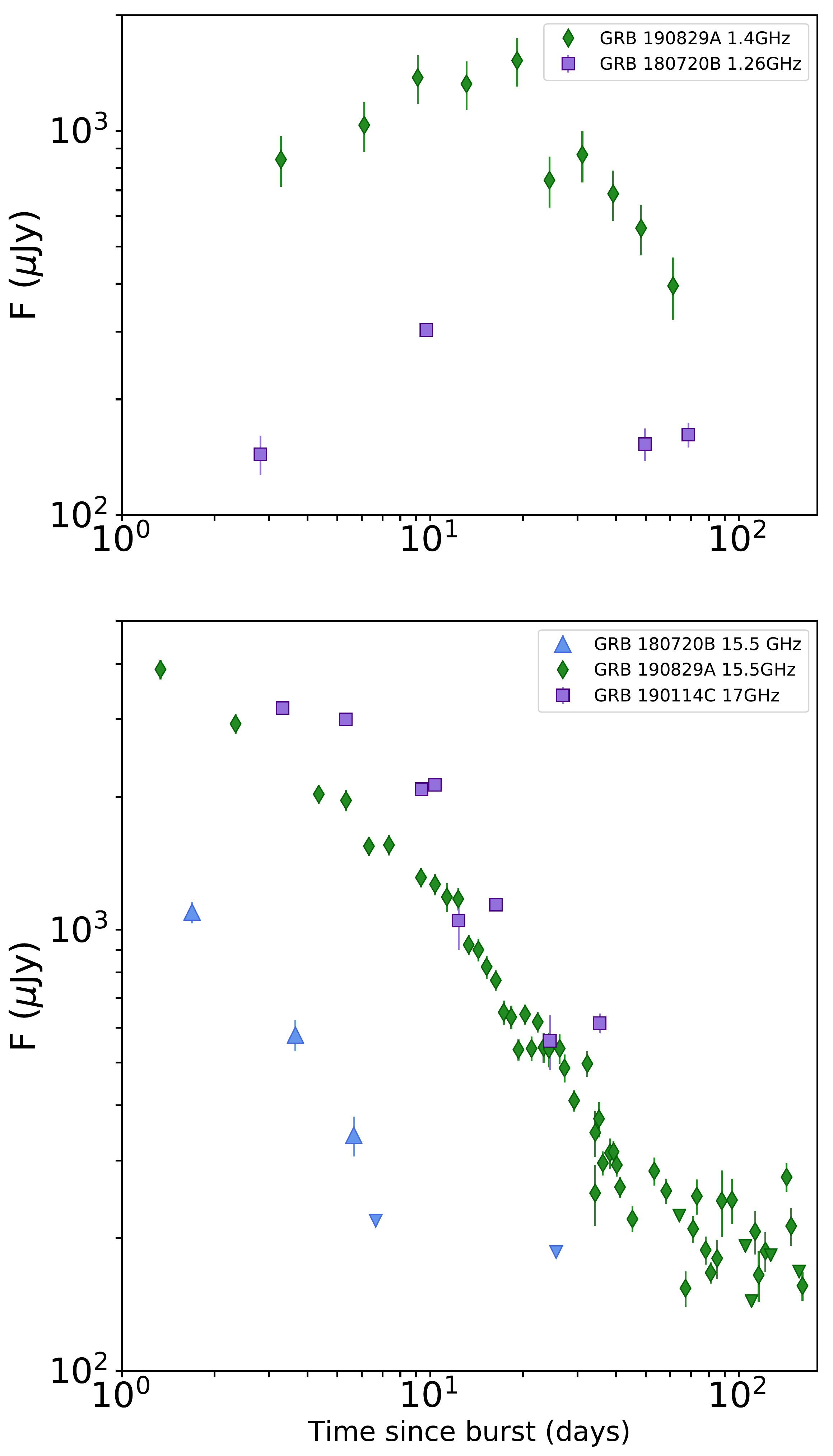}
    \caption{Light curves from the MAGIC GRB 190114C, H.E.S.S. GRB 180720B and H.E.S.S. GRB 190829A for similar frequencies. \textit{Upper panel}: shows the low frequencies ($\sim$1\,GHz) light curves for GRB 190114C \citep[from uGRMT - the squares, and from MeerKAT - the stars,][]{2019arXiv191109719M, 2019Natur.575..459M} and 190829A (this work). \textit{Lower panel}: shows the high frequency ($\sim$16\,GHz) light curves for all three VHE GRBs \citep{2019arXiv191109719M}}
    \label{fig:GRB_190114C}
\end{figure}

Figure \ref{fig:GRB_190114C} shows a comparison between H.E.S.S. GRBs 190829A and 180720B, and MAGIC GRB 190114C. The upper panel shows the light curve of GRB 190114C observed with upgraded Giant Metrewave Radio Telescope and MeerKAT, along with our GRB 190829A MeerKAT light curve, both in the 1.3 GHz band \citep{2019arXiv191109719M, 2019Natur.575..459M}. The lower panel shows the AMI-LA H.E.S.S. GRB 190829A and H.E.S.S. GRB 180720B light curves, and Australia Telescope Compact Array 17\,GHz data for MAGIC GRB 190114C. We note that the GRB 190114C fluxes have not been host galaxy-corrected \citep{2019arXiv191109719M, 2019GCN.23760....1T}. 

The low frequency light curve for GRB 190114C from \citet{2019arXiv191109719M, 2019Natur.575..455M}, in the upper panel of Figure \ref{fig:GRB_190114C}, shows a rise and decay over four epochs with a peak at about ten days. The light curve is qualitatively similar to the GRB 190829A low frequency light curve, both from the FS, however unlike for GRB 190829A, GRB 190114C is shown to have a steeper rise and has an optically thick spectrum too. The higher frequency light curve for GRB 190114C can be described by a broken power law with a break also at 10 days. The slow decay, from early times, which is also seen in the optical and near infra-red, is also attributed to a FS component \citep{2019arXiv191109719M}. Only GRB 190829A shows two separate shocks, the afterglows from GRB 190114C and 180720B are modelled in terms of the FS only. However, this may be due to limited light curve sampling of the latter two GRBs compared to the GRB 190829A radio afterglow, which makes it difficult to model their radio radio with two components.

\subsection{VHE GRBs: comparison with a flux-limited sample}

\begin{figure*}
    \centering
    \includegraphics[width=0.8\textwidth]{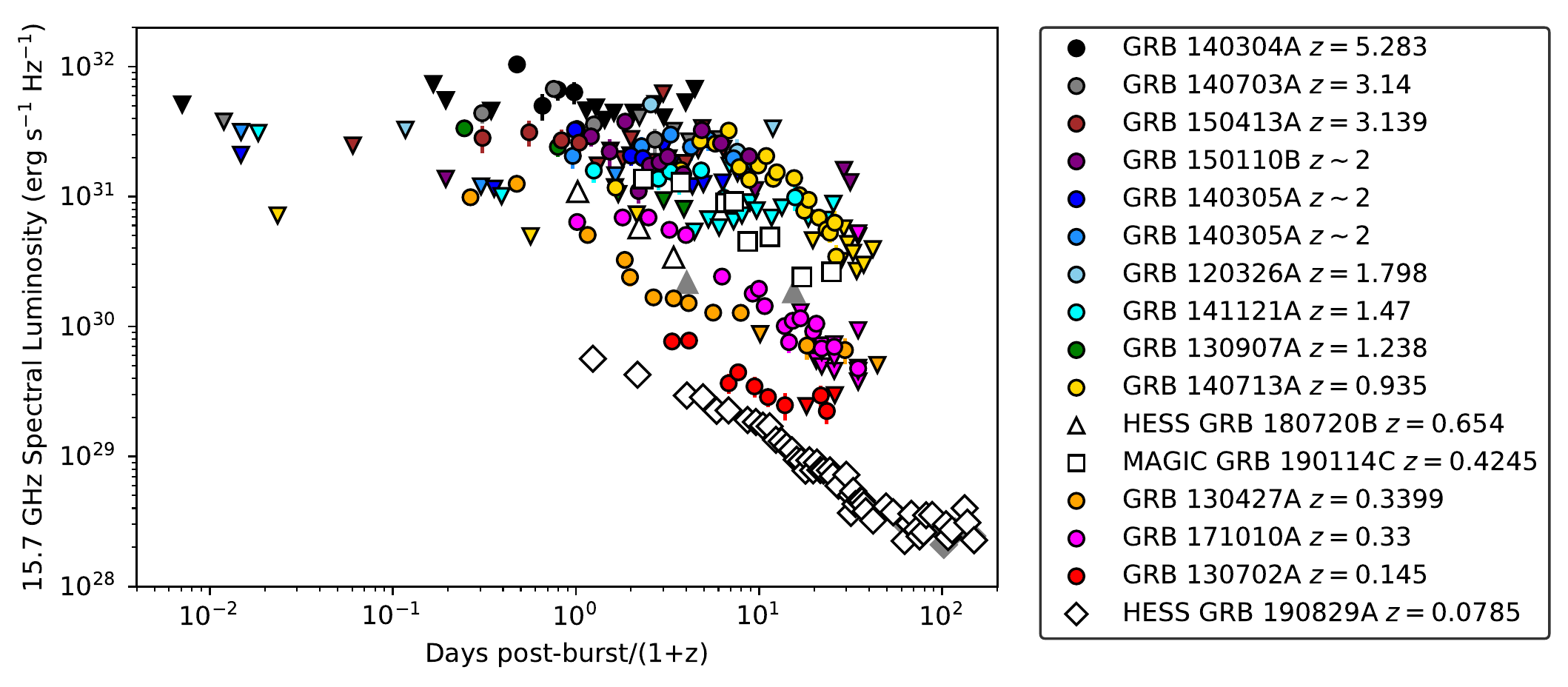}
    \caption{Comparison of data from MAGIC GRB 190114C, H.E.S.S. GRB 180720B and H.E.S.S. GRB 190829A at frequencies around 16\,GHz for the VHE GRBs with a sample of the GRB population as detected by AMI-LA \citep{Anderson2017, 2019MNRAS.486.2721B, 2019arXiv191109719M}. The GRBs from the AMI-LA catalogue are given as multi-coloured circles for detections and down-facing triangles for 4$\sigma$ upper limits. For GRBs without known redshifts, z = 2 is used. The VHE GRBs are the white symbols with black edges and grey up-facing triangles for 3$\sigma$ upper limits. }
    \label{fig:AMI GRB catalogue}
\end{figure*}

We compare the spectral radio luminosities of the three GRBs with each other and a sensitivity limited sample of radio detected GRB population using the AMI-LA GRB catalogue \citep{Anderson2017, 2019MNRAS.486.2721B}. The spectral luminosities are calculated using $L = 4\pi F_{\nu}D_{L}^2(z+1)^{\alpha-\beta-1}$ where $F_{\nu}$ is the measured flux, $D_{L}$ is the luminosity distance, $\alpha$ and $\beta$ are the temporal and spectral indices, set to 0 and $\frac{1}{3}$, respectively, according to \citet{2012ApJ...746..156C}. We assume a flat $\Lambda$CDM Universe with $H_0=68$km\,Mpc\textsuperscript{-1}s\textsuperscript{-1} and $\Omega_M = 0.3$. Figure \ref{fig:AMI GRB catalogue} shows a direct comparison between the VHE GRBs and the AMI-LA GRB catalogue.

The radio spectral luminosity of GRB 190829A is two orders of magnitude lower than the two other VHE GRBs, in addition to having the lowest luminosity in the entire sample. Taken alone it could be suggested that VHE GRBs are from a low-radio-luminosity sub-sample. However, GRB 190829A is only marginally fainter than GRB 130702A, which has no VHE emission detected above a few GeV \citep{2013GCN.14971....1C}. In addition, the luminosities for GRB 190114C sits firmly within the bulk of the GRB sample with GRB 180720B bordering the low end of the group but still two orders of magnitude brighter than GRB 190829A. When considering all three VHE-GRBs there appears to be no difference in the radio luminosities between the VHE and non-VHE GRBs. Despite the range of luminosities of the VHE GRBs in Figure \ref{fig:AMI GRB catalogue}, none stand out with respect to this sensitivity limited sample of radio detected GRBs.

A comparison can also be made using the isotropic energy (E\textsubscript{ISO}) of the three events and that of the rest of the population. Konus Wind \citep{Aptekar1995} observations show GRBs 190114C, 190829A, and 180720B have $\textrm{E}_{\textrm{ISO}} = 3\times10^{53}\textrm{erg}, 2\times10^{50}\textrm{erg}$, and $6\times10^{53}\textrm{erg}$, respectively \citep{2019GCN.23737....1F, 2019GCN.25660, 2018GCN.22973....1S}. The E\textsubscript{ISO} for GRB 190829A is three orders of magnitude fainter than the other two VHE GRBs making GRB 190829A only detectable because it was very nearby, as shown by figure 1 of \citet{2014ApJ...781...37P}. GRBs with lower isotropic energies have been detected, such as GRB 980425, 060218, 100316D, which all have E\textsubscript{ISO} in the range of $10^{48-49}$erg but all are at redshifts lower than $\sim$0.1 (z = 0.0085, 0.033, 0.059, respectively) like that of GRB 190829A (\citealt{2014ApJ...781...37P} and references therein). 

We have shown that these VHE GRBs with radio detections are similar to other GRBs with observed radio afterglows, and as a result one might expect more VHE counterparts to have been detected. Above z = 1, detections of GRBs by Cherenkov facilities are far less likely, due to the Universe's high opacity to VHE gamma-rays  \citep{2017A&A...606A..59H}. However, there are many GRBs, with z<1, that have been detected without VHE counterparts, such as GRB 100621A \citep{2014A&A...565A..16H}. It is possible that the range of luminosities of GRBs at VHE energies are such that H.E.S.S. observations are sensitivity limited and so many VHE counterparts are undetectable with the current instrumentation. 

\section{Conclusions}

We have presented nearly 200 days of radio and X-ray observations of H.E.S.S. GRB 190829A. MeerKAT 1.3 GHz and \textit{Swift}-XRT light curves appear to be dominated by forward shock emission while the AMI-LA data at 15.5 GHz appear instead to be dominated by a reverse shock up to at least 50 days. We show that neither shock component significantly contributes to the flux at the other radio frequency. In addition to emission from the GRB we also see the host galaxy in both radio radio bands, at 1.3\,GHz we see the galaxy spatially resolved from the GRB position. Applying a standard fireball model to the data, it can be concluded that the circumburst medium is homogeneous. 

We have also presented previously unpublished AMI-LA observations of GRB 180720B, the first GRB with a VHE detection. Comparison between these two H.E.S.S GRBs, GRB 190829A and GRB 180720B, MAGIC GRB 190114C, and a sample of GRBs without detected VHE emission show no significant differences. This is consistent with the VHE GRBs being drawn from the same parent population as the other radio-detected long GRBs.

\section*{Data availability statement}

The data underlying this article are available in the article with two exceptions. The data for Figure \ref{fig:XRT} is found at \href{https://www.swift.ac.uk/xrt\_curves/00922968/}{https://www.swift.ac.uk/xrt\_curves/00922968/}. The data for Figure \ref{fig:AMI GRB catalogue} was provided by G.E. Anderson by permission. Said data can be shared on request with permission of G.E. Anderson.

\section*{Acknowledgements}
The authors would like the thank the anonymous referee for their helpful comments. L. Rhodes and J.S. Bright acknowledge the support given by the Science and Technology Facilities Council through an STFC studentship. I. M. Monageng acknowledges support from the National Research Foundation of South Africa. G. E. Anderson is the recipient of an Australian Research Council Discovery Early Career Researcher Award (project number DE180100346) funded by the Australian Government. M. F. Bietenholz was supported by both the National Sciences and Engineering Research Council of Canada and the National Research Foundation of South Africa. The work of M. B\"ottcher is supported by the South African Research Chairs Initiative (grant no. 64789) of the Department of Science and Innovation and the National Research Foundation\footnote{Any opinion, finding and conclusion or recommendation expressed in this material is that of the authors and the NRF does not accept any liability in this regard.} of South Africa. S.E. Motta acknowledges the Violette and Samuel Glasstone Research Fellowship programme for financial support. S.E. Motta and D.R.A. Williams acknowledge support by the Oxford Centre for Astrophysical Surveys, which is funded through generous support from the Hintze Family Charitable Foundation. P.A. Woudt acknowledges the University of Cape Town and the National Research Foundation for financial support. This work made use of data supplied by the UK Swift Science Data Centre at the University of Leicester. The authors would like to thank the anonymous referee for their helpful comments. We also thank SARAO for the approval of the DDT MeerKAT requests and the schedulers for their time in making the observations. The MeerKAT telescope is operated by the South African Radio Astronomy Observatory, which is a facility of the National Research Foundation, an agency of the Department of Science and Innovation. We thank the Mullard Radio Astronomy Observatory staff for scheduling and carrying out the AMI-LA observations. The AMI telescope is supported by the European Research Council under grant ERC-2012-StG-307215 LODESTONE, the UK Science and Technology Facilities Council, and the University of Cambridge.  This research has made use of NASA's Astrophysics Data System, and the Python packages \textsc{numpy} \citep{5725236} and \textsc{matplotlib} \citep{4160265}.



\bibliographystyle{mnras}
\bibliography{bibliography.bib}


\appendix

\section{MCMC results}
\label{section:appendixB}

Here we show the posterior probability distributions of the fitted parameters from fitting equation \ref{eq:bpl} to the data.  
\vfill\null

\begin{figure}
    \centering
    \includegraphics[width=\columnwidth]{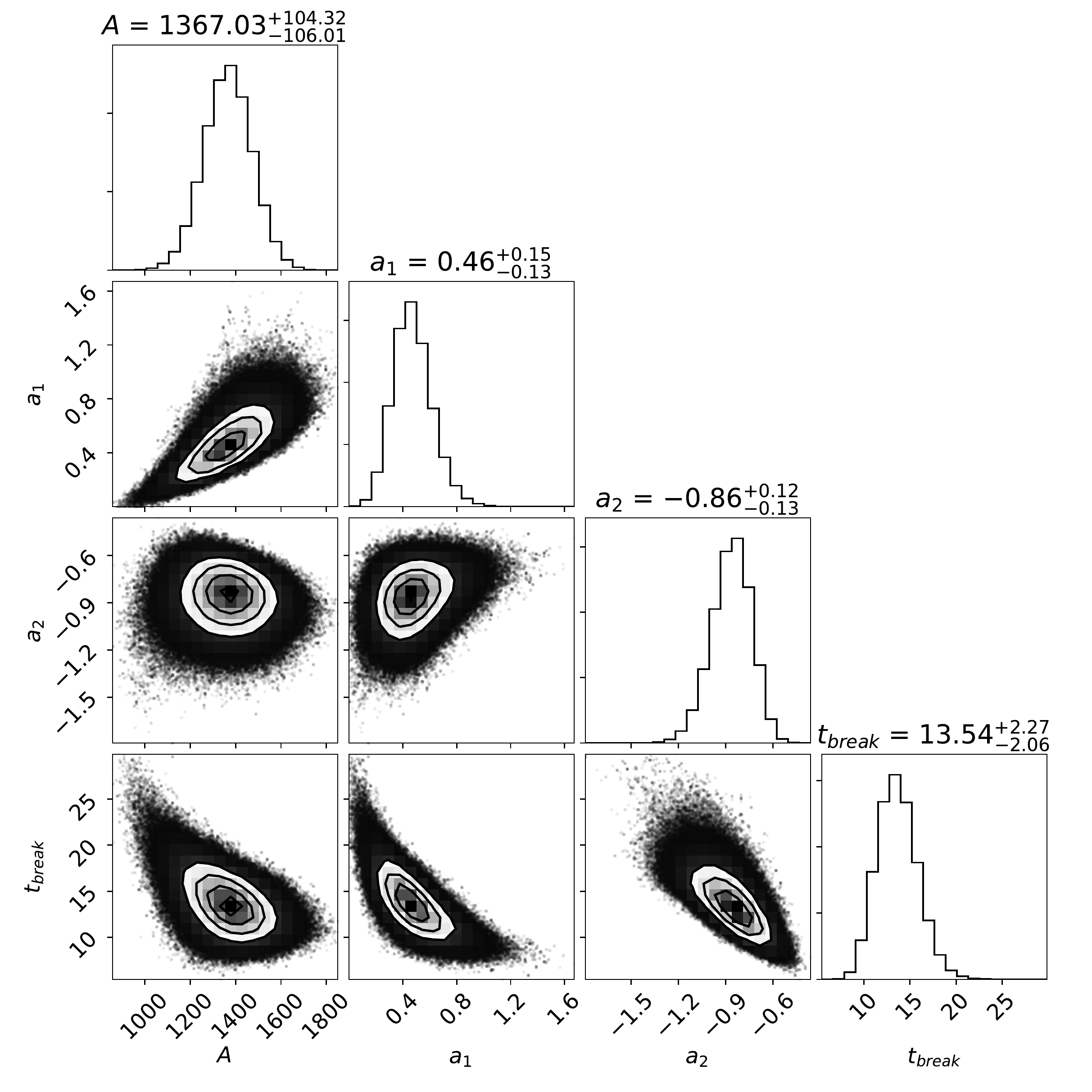}
    \caption{Corner plots from MCMC fitting code for fitting a broken power law to the 1.3\,GHz MeerKAT data}
    \label{fig:corner MeerKAT}
\end{figure}

\begin{figure}
    \centering
    \includegraphics[width=\columnwidth]{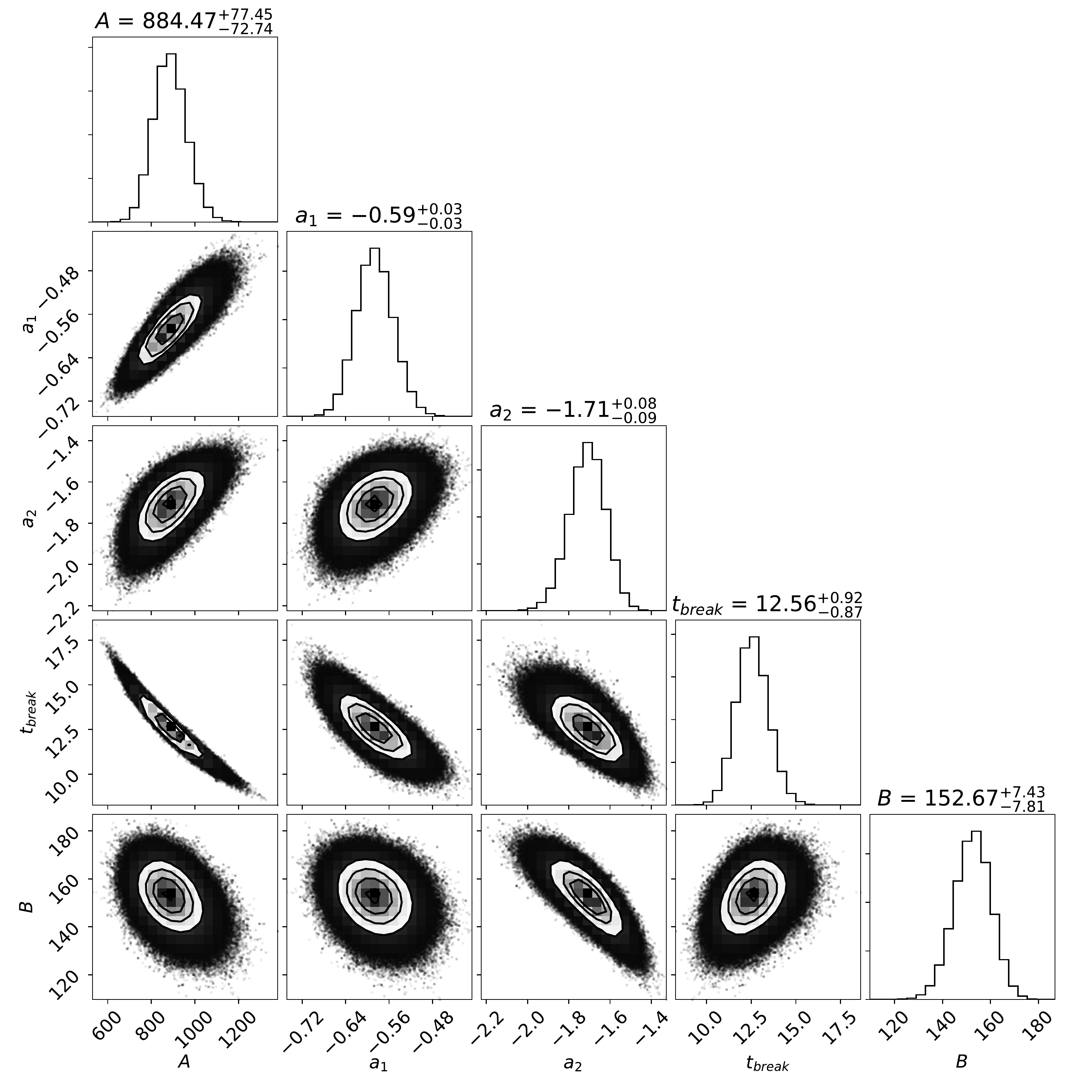}
    \caption{Corner plots from MCMC fitting code to fit a broken power law with a constant component to 15.5GHz AMI-LA data}
    \label{fig:corner AMI}
\end{figure}


\bsp	
\label{lastpage}
\end{document}